\documentclass[aps,pra,superscriptaddress,twocolumn,showpacs]{revtex4}
\usepackage{amsmath}
\usepackage{amssymb}
\usepackage{bm}
\usepackage{epsfig}
\usepackage{graphicx}
\usepackage{color}
\usepackage{dcolumn}
\usepackage{bm}

\begin{document}
\title{Temporal oscillations of light transmission through dielectric microparticles subjected to optically induced motion}

\author{Almas F. Sadreev}
\affiliation{Kirensky Institute of Physics, 660036 Krasnoyarsk, Russia}
\author{E. Ya. Sherman}
\affiliation{Department of Physical Chemistry, Universidad del Pa\'is Vasco UPV-EHU,
48080 Bilbao, Spain}
\affiliation{IKERBASQUE Basque Foundation for Science, Bilbao, Spain}

\date{\today}

\begin{abstract}
We consider light-induced binding and motion of dielectric microparticles in an optical
waveguide that gives rise to a back-action effect such as
light transmission oscillating with time. Modeling the particles by dielectric slabs allows 
us to solve the problem analytically and obtain 
a rich variety of dynamical regimes both for Newtonian and damped motion.
This variety is clearly reflected in temporal oscillations of the light transmission.
The characteristic frequencies of the oscillations are within the ultrasound range of the order of
$10^{5}$ Hz for micron size particles and
injected power of the order of 100 mW. 
In addition, we  consider driven by propagating  light dynamics of a dielectric particle inside a Fabry-Perot
resonator.
These phenomena pave a way for optical driving and monitoring of motion of particles in waveguides
and resonators.
\end{abstract}
\maketitle

\section{Introduction}
The response of a microscopic dielectric object to an optical 
field can profoundly affect its motion. A classical example
of this influence is an optical trap, which can hold a particle in a tightly
focused light beam \cite{Ashkin}.
Optical fields can also be used to arrange,
guide or detect particles in appropriate light-field geometries
\cite{Burns,Ogura,Yang,Brzobohaty}. Optical forces are ideally suited
for manipulating microparticles in various systems, which are
characterized by length scales ranging from hundreds of
nanometers to hundreds of micrometers, forces ranging
from femto- to nanonewton, and time scales ranging upward from a microsecond \cite{Grier}.
Transportation of particles of various sizes by light
is of an immense  growing interest caused by many potential applications \cite{Ogura,Vahala,Barker}.

\begin{figure}
\includegraphics*[width=8cm,clip=]{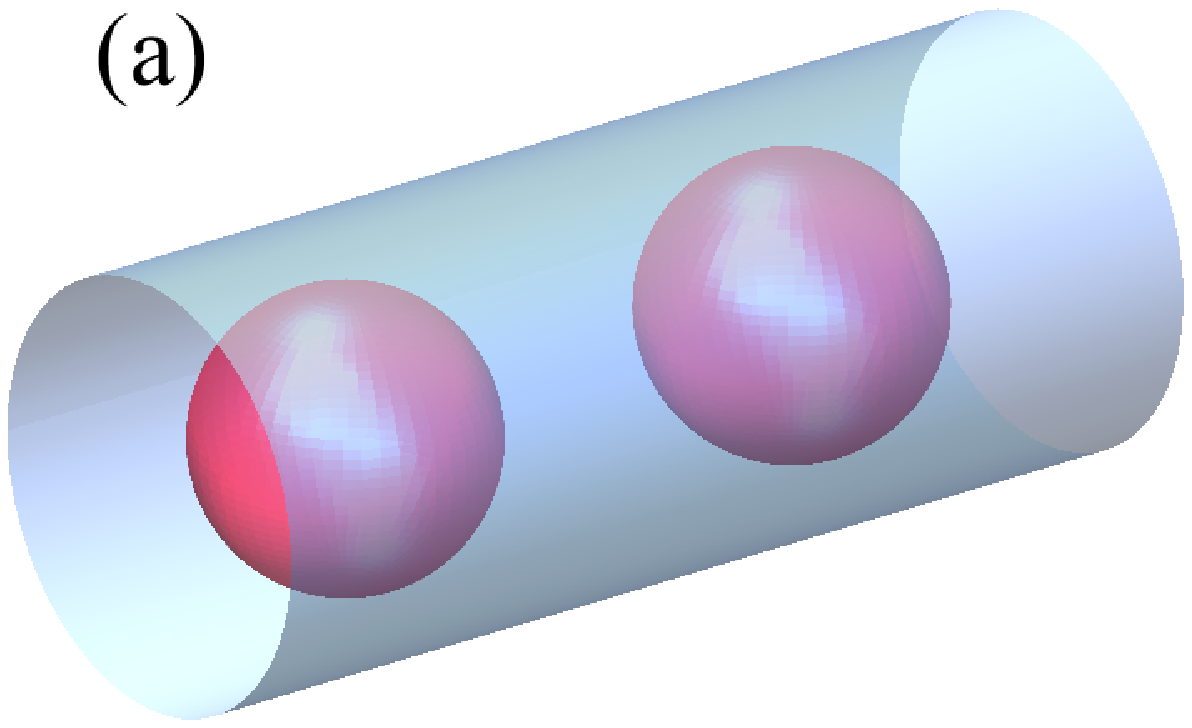}\\
\includegraphics*[width=7cm,clip=]{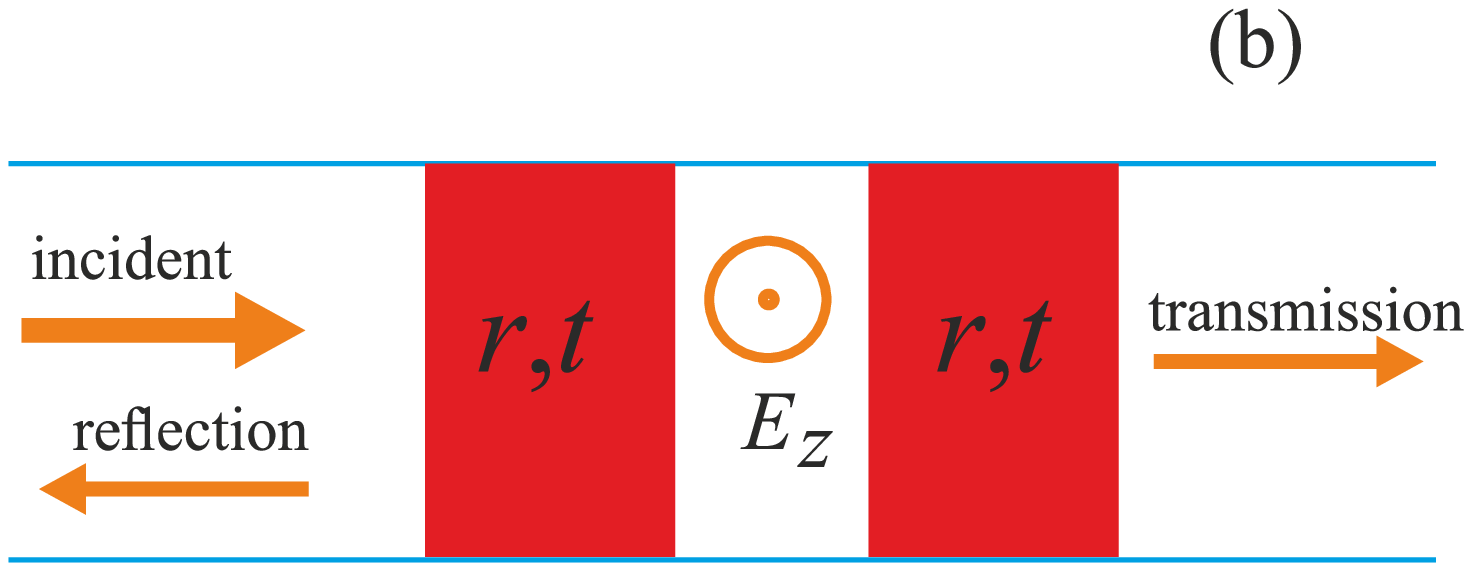}
\caption{ (a) Two particles inside a waveguide.
(b) Model of two identical dielectric slabs in waveguide for the TE transmission. Coefficients $r$ and $t$ characterize reflection and transmission 
of each slab and $E_{z}$ corresponds to the direction of electric field inside the waveguide.}
\label{figure:model}
\end{figure}

Manipulation of dielectric objects of a submicron size requires a
strong optical confinement and high intensities than can be provided
by diffraction-limited systems \cite{Yang}.
In order to overcome these limitations it was proposed to use
subwave length liquid-core slot waveguides \cite{Almeida}, fiber or photonic crystal
(PhC) waveguides and cavities \cite{Barker,Barth,Hu}.
The technique simultaneously makes use of near-field optical forces to
confine particles inside the waveguide and scattering/absorption forces to
transport it. The ability of the slot or the PhC waveguides to condense the accessible
electromagnetic energy to spatial scales as small as 60 nm also allows researchers to
overcome the fundamental diffraction problem.
However the consequence is that the cavity mode is strongly perturbed by
the presence of a particle in its vicinity making standard
PhC cavities unsuitable for noticeable back action effects.
A clear evidence of the back action between a resonant field in a
photonic crystal cavity and a single dielectric nanoparticle through
the optical gradient forces was presented in Refs.
\cite{Barth,Juan,Roichman,Descharmes}. As a result, the motion of the particles
can considerably modify the light propagation.

The aim of the present paper is to study the time dependent back action effect
for light propagation in a waveguide including a few
dielectric microparticles with sizes comparable with the light wavelength, similar to an example
shown in Fig. \ref{figure:model}(a). An analogous problem was considered by Kar\'{a}sek {\it et al.}
\cite{Karasek} who numerically studied by the coupled dipole method a
longitudinal optical binding between two microparticles in a Bessel beam.

There are several aspects tremendously complicating the consideration of spherical particles.
(i) Spheres give rise to the
problem of calculation of electromagnetic (EM) fields of both polarizations especially in the near-field zone. This
problem  can be solved only numerically by expanding of the waveguide propagating solutions over
vector spherical functions and using the Lorenz-Mie theory \cite{Stratton,Lock}.
(ii) For the scattering the Mie resonances could play important role for the dielectric
spheres of high refractive index.
(iii) All translational and rotational degrees of freedom
are to be included in the dynamics of each particle.

In the present paper we model the particles with dielectric slabs
inserted in a directional waveguide of a square cross-section $d\times d$
as shown in Fig. \ref{figure:model} (b). We take the perpendicular dimensions of the slabs very close to this cross-section.
This allows us to consider only {one-dimensional} motion of particles
and treat the problem analytically. This approach was applied for calculation of
optical forces on dielectric particles in one-dimensional optical
lattices \cite{Asboth,Xuereb,Sonnleitner} by using the transfer matrix \cite{Markos}.
This model of a classical optomechanical system \cite{Aspelmeyer} preserves all qualitative
features of the initial problem as it can be described by the transfer matrix and predicts
the important result of temporal oscillations of light transmittance caused by light-induced motion.

This paper is organized as follows. In Sec. II we remind the reader
the formulas for the light pressure on a single particle in a waveguide.
In Sec. III we formulate the model and consider motion of a single particle
in the presence of a static ``scattering center'' inserted in the waveguide. In Sec.
IV we investigate regimes of motion of two mobile particles. Section V presents the
results for motion of a single particle inside a Fabry-Perot resonator. Conclusions and
discussion of the results are given in Sec. VI.

\section{Forces on a dielectric slab in a waveguide}

Motion of a particle in a vacuum- or air-filled waveguide is governed by the EM force $\mathbf{F}$
defined by the stress-tensor $T_{\alpha\beta}$ integrated over the surface elements $dS_{\beta}$ \cite{LL,Pendry}
\begin{eqnarray}\label{force}
&&F_{\alpha}=\int T_{\alpha\beta}dS_{\beta}, \\
&&T_{\alpha\beta}=
    \frac{1}{4\pi}E_{\alpha}E_{\beta}^{*}-\frac{1}{8\pi}\delta_{\alpha\beta}
|\mathbf{E}|^2+ \nonumber \\
&&\qquad\frac{1}{4\pi}H_{\alpha}H_{\beta}^{*}-\frac{1}{8\pi}\delta_{\alpha\beta}
   |\mathbf{H}|^2, \nonumber
\end{eqnarray}
where $\alpha$ and $\beta$ are the Cartesian indices. 
We concentrate on the basic propagating mode $\mbox{TE}_{10}$  having the following solution \cite{Jackson}
\begin{eqnarray}\label{TE}
H_x&=&H_{0}\psi(x)\cos\frac{\pi y}{d},\nonumber\\
H_y&=&-\frac{ikd}{\pi}H_{0}\psi(x)\sin\frac{\pi y}{d},\\
E_z&=&\frac{i\omega d}{\pi}H_{0}\psi(x)\sin\frac{\pi y}{d},\nonumber
\end{eqnarray}
where
\begin{equation}\label{omega}
    \omega^2=\frac{\pi^2}{d^2}+k^2,
\end{equation}
$H_{0}$ is the field amplitude, $\psi(x)=e^{ikx}$ in the uniform waveguide, and the speed of light $c\equiv 1$.

\begin{figure}
\includegraphics*[width=7cm,clip=]{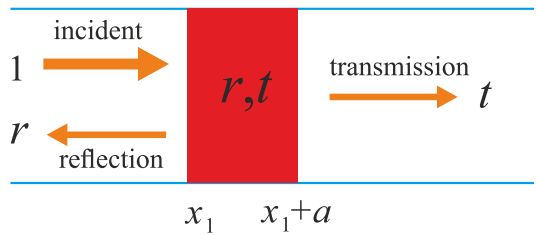}
\caption{Transmission and reflection through a single slab.}
\label{figure:singleslab}
\end{figure}

To describe the EM field we need to know the scattering properties of each
slab specified by the reflection and
transmission coefficients $r, t$ which can be expressed with the transfer matrix
\textbf{$\mathbf{M}$} \cite{Markos}
\begin{eqnarray}\label{M}
M_{11}&=&\cos(qa)+\displaystyle{\frac{i}{2}\left[\frac{q}{k}+\frac{k}{q}\right]
\sin(qa)},\nonumber\\
M_{12}&=&\displaystyle{\frac{i}{2}\left[\frac{q}{k}-\frac{k}{q}\right]\sin(qa)},~
M_{22}=M_{11}^{*}, ~M_{21}=M_{12}^{*},\nonumber\\
t&=&\displaystyle{\frac{1}{M_{22}}},~~ r=\displaystyle{\frac{M_{12}}{M_{22}}},
\end{eqnarray}
where $a$ is  the slab thickness. 
Here $q$ is wave vector component along the $x$-axis given by:
\begin{equation}\label{q}
q^2=\epsilon k^2+(\epsilon-1)\frac{\pi^2}{d^{2}},
\end{equation}
where $\epsilon$ is the dielectric constant of the slabs shown in Fig. \ref{figure:singleslab}.

Let us consider force acting on such a slab. Its presence in the waveguide 
modifies the components of the electromagnetic field in the $\mbox{TE}_{10}$ mode (\ref{TE})
as:
\begin{equation}\label{psi}
\frac{H_x}{H_0}=\cos\frac{\pi y}{d}\left\{
\begin{array}{l}
e^{ik(x-x_1)}+re^{-ik(x-x_1)},\quad x<x_1,\cr\cr
te^{ik(x-x_1-a)},\quad  x>x_1+a,\cr
\end{array}
\right.
\end{equation}
\begin{equation}
\frac{H_y}{H_{0}}=-\frac{ikd}{\pi}\sin\frac{\pi y}{d}
\left\{\begin{array}{l}
e^{ik(x-x_1)}-re^{-ik(x-x_1)},\quad x<x_1,\cr\cr
te^{ik(x-x_1-a)},\quad x>x_1+a,\cr
\end{array}\right.
\end{equation}
\begin{equation}
\frac{E_z}{H_0} =\frac{i\omega d}{\pi}\sin\frac{\pi y}{d}
\left\{\begin{array}{l}
e^{ik(x-x_1)}+re^{-ik(x-x_1)},\quad x<x_1,\cr\cr
te^{ik(x-x_1-a)},\quad x>x_1+a.\cr
\end{array}\right.
\end{equation}
Substituting  these solutions into Eq. (\ref{force}) we obtain the light pressure
\begin{equation}\label{force1}
    P=P_0(1+|r|^2-|t|^2)=2P_0|r|^2,
\end{equation}
where
\begin{equation}\label{F0}
    P_0=\frac{H_0^2}{8\pi}\left(\frac{kd}{\pi}\right)^2.
\end{equation}

\section{Dynamics of a single particle in the  presence of a scattering center}

\begin{figure}[htbp]
\centering
\includegraphics*[width=7cm,clip=]{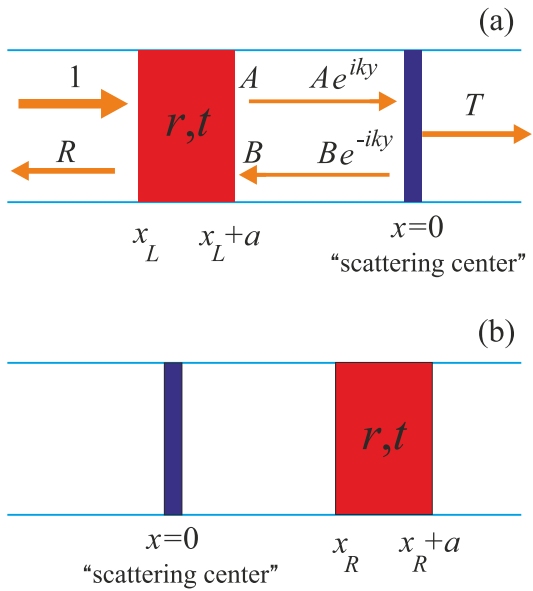}
\caption{Immobile scattering center-related geometries. (a) Movable slab is on the left from the scattering 
center and $y=-(x_{L}+a)$. In this geometry we consider two types of the static  elements: 
one optically equivalent to the movable slab and an ideal mirror. $A$ and $B$ are the field amplitudes. 
(b) Movable slab is on the right
from the scattering center, which here is optically equivalent to the slab.}
\label{figure:scatteringcenter}
\end{figure}

The situation described in the previous section changes dramatically 
if another element, besides the mobile dielectric particle is inserted in the waveguide.
In particular, we insert at $x=0$ an immobile particle (``scattering center'') characterized by light transmission and
reflection coefficients, as shown in Fig. \ref{figure:scatteringcenter}.
Then the solution for the EM field and therefore, the force acting on the mobile particle
become dependent on its distance to the center and cause various regimes of the slab motion.
Here we concentrate on this motion driven by the optical force $F_j(x_j)$ and the corresponding 
potential $U_j(x_j)$ described by equation
\begin{equation}\label{dynL}
    m\ddot{x}_{j}+6\pi\eta d\dot{x}_{j}=F_j(x_j)=-\displaystyle{\frac{dU_j}{dx_j}},
\end{equation}
resulting, as we will show, in the time oscillations of the light transmittance through the waveguide.
Index $j=L, R$ enumerates the particle positioned on the left or on the right
from the immobile center, $m=\rho ad^2$ is the particle mass ($\rho$ is the material density), and
$6\pi\eta d$ is the linear drag coefficient for a particle in a medium of viscosity
$\eta$ \cite{Praveen,coefficient}.
In what follows we choose the dielectric constant of the slab $\epsilon=4$ (glass),
its width $a=d/2$, and zero initial velocity. We neglect imaginary part of the dielectric
constant and corresponding contribution into the optical force due to its smallness in the visible light frequency domain 
\cite{Kitamura}.

We begin with the realization shown in Fig. \ref{figure:scatteringcenter}(a). 
Similar to Ref. \cite{Xuereb}, we write the equation for
the ingoing and outgoing amplitudes of waves $\psi(x)$ describing
the EM field components in each region of the waveguide (Fig. \ref{figure:scatteringcenter}):
\begin{equation}
\label{AB}
\left[\begin{array}{c}
  A\\ B\end{array}\right]=\mathbf{M}\left[\begin{array}{c}1\\  R\end{array}\right],\quad
  \left[\begin{array}{c}T\\0\end{array}\right]=\mathbf{M}\left[\begin{array}{c}
  Ae^{iky}\\ Be^{-iky} \end{array}\right],
\end{equation}
where we assumed optical equivalence of the scattering center and the movable slab, 
$y=-(x_{L}+a)$ is the distance between the particles, and the matrix $\mathbf{M}$ is given
by Eq. (\ref{M}).
The total transmission and reflection amplitudes can be expressed as \cite{Markos,BW}
\begin{equation}\label{RT}
  R=r+\displaystyle{\frac{t^2re^{2iky}}{1-r^2e^{2iky}}},\qquad
T=\frac{t^2e^{iky}}{1-r^2e^{2iky}}.
\end{equation}
Substituting the solution of Eq. (\ref{AB}) into Eq. (\ref{force}) we find the
forces acting on the slab ($F_{L}$) and the scattering center ($F_{\rm SC}$) as 
\begin{eqnarray}\label{forces}
&&F_{L}(y)=P_{0}d^{2}\left[1+|R|^2-|A|^2-|B|^2\right],\nonumber\\
&&F_{\rm SC}(y)=P_{0}d^{2}\left[|A|^2+|B|^2-|T|^2\right]\nonumber\\
&&=-F_{L}(y)+2P_{0}d^{2}|R|^{2},
\end{eqnarray}
respectively. The forces depend only on the distance $y$.

\begin{figure}[htbp]
\begin{centering}
\includegraphics*[width=5cm,clip=]{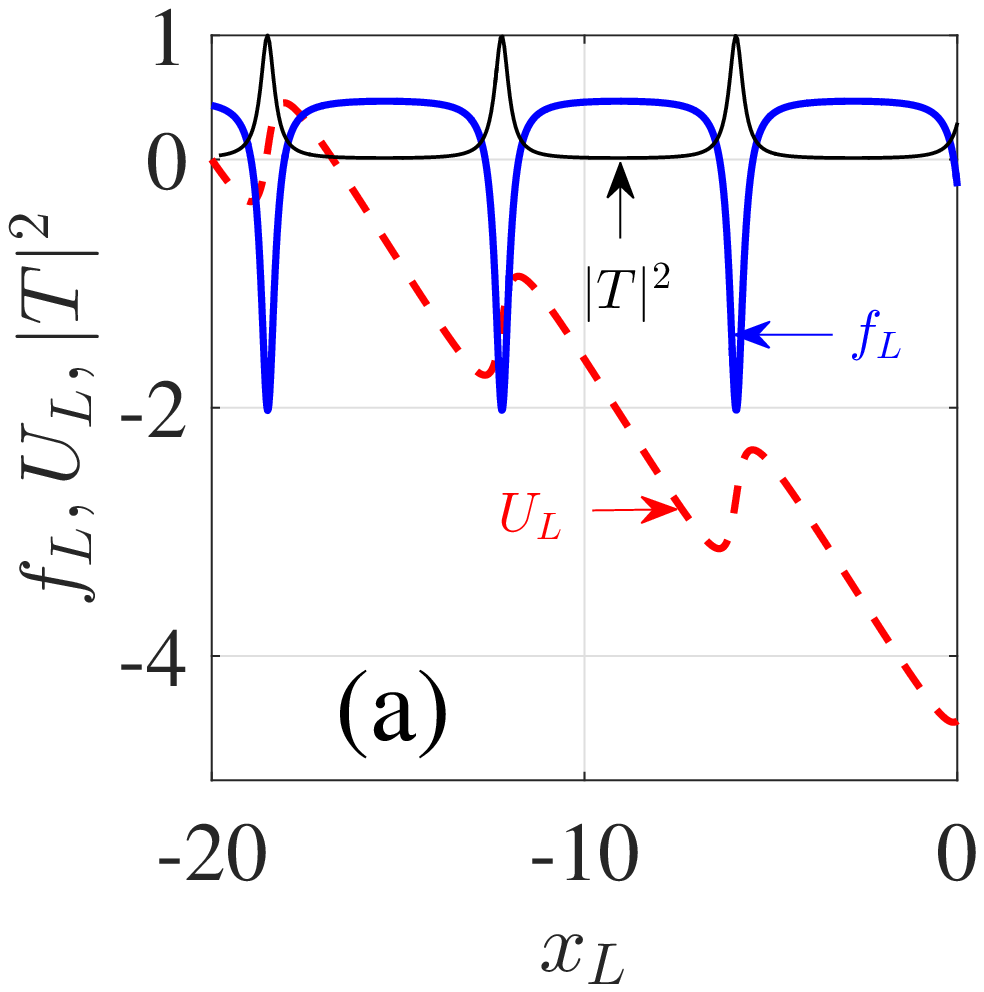}\\
\includegraphics*[width=5.4cm,clip=]{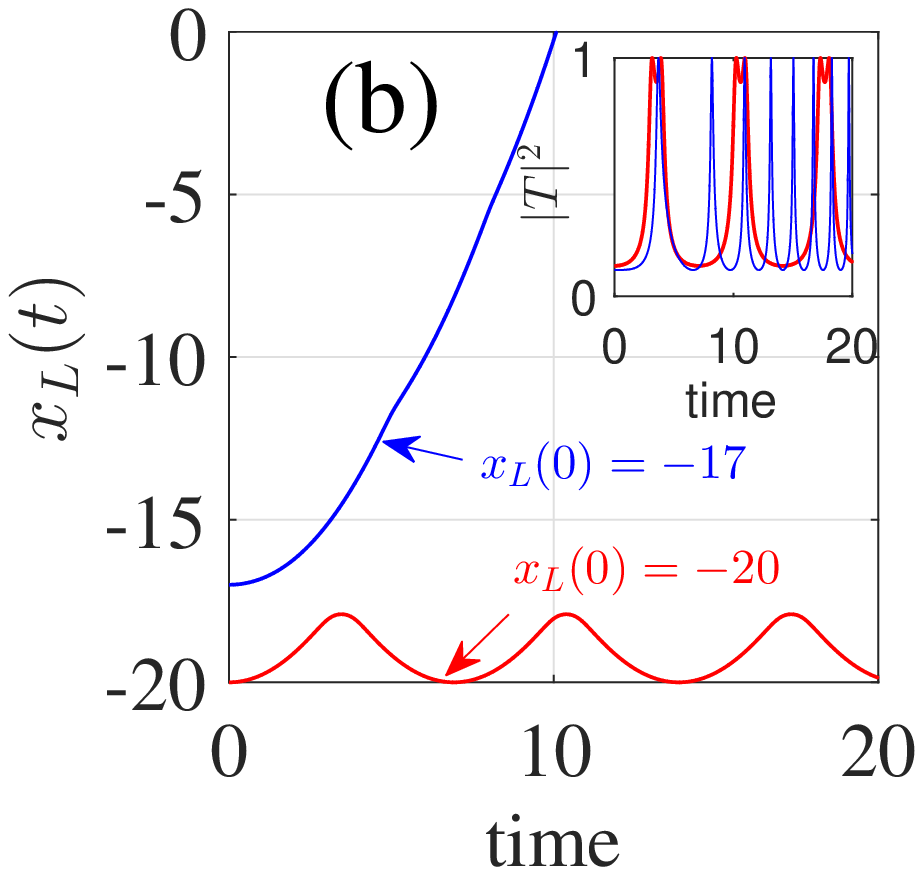}\\
\includegraphics*[width=5cm,clip=]{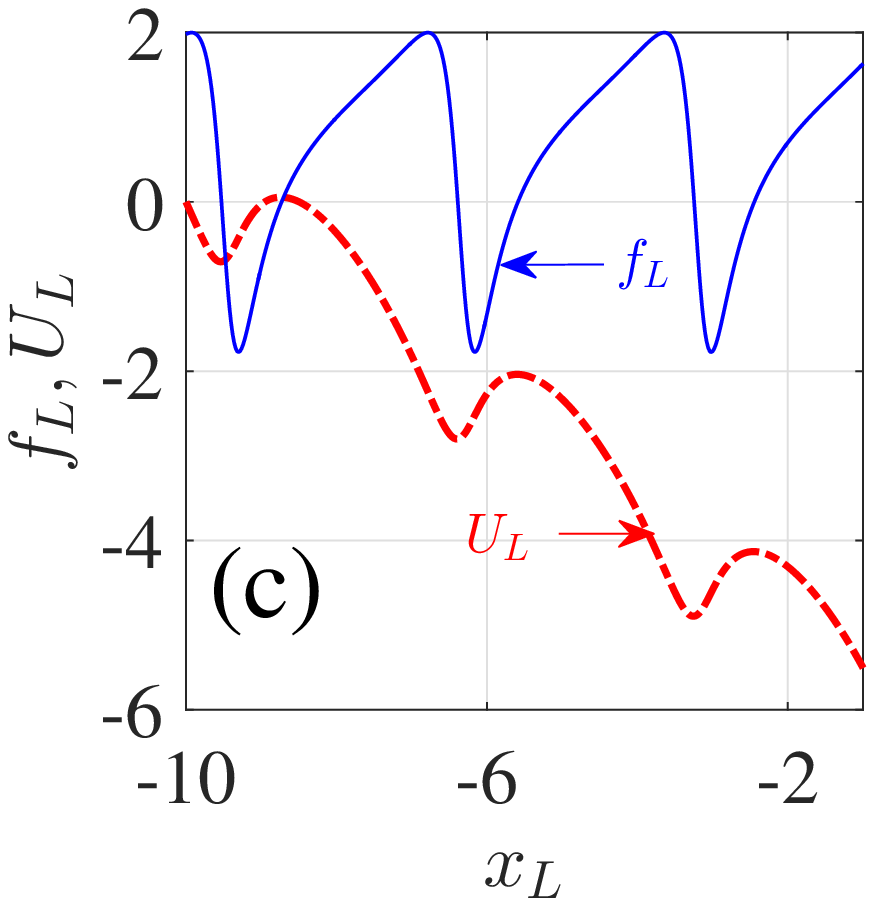}\\
\includegraphics*[width=5cm,clip=]{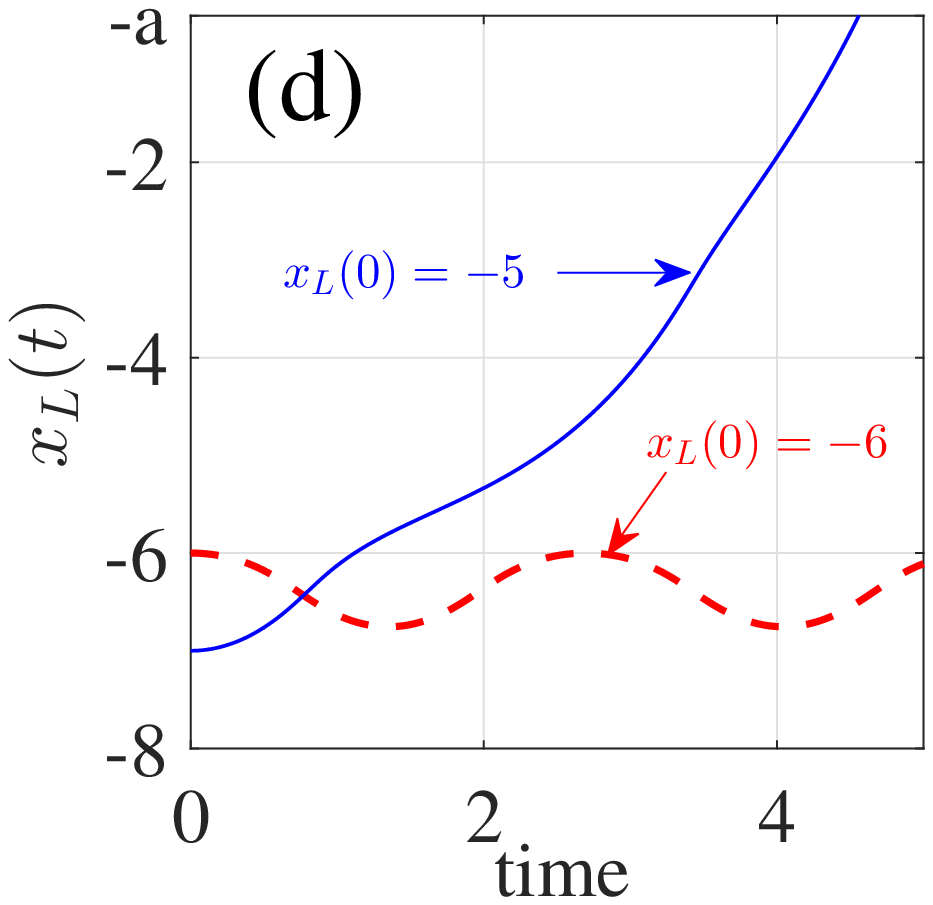}
\end{centering}
\caption{(a) The optical force (\ref{forces}) (solid line), the corresponding potential (dot line), 
and the light transmission (dash-dot line) for the first (left)
moving particle and the immobile (right) particle positioned at $x_R=0$.
(b) The corresponding time evolution of the left particle position and light transmission
for two choices of initial positions.
Figures (c) and (d) correspond to an immobile mirror with at the position $x_R=0$.
The parameters of the slab are $a=d/2, \epsilon=4, k=1/2$. The evolution in (a)-(d)
is considered frictionless with $\gamma=0$. The coordinates, forces, and time are given in the units of $d$,
$P_{0}d^{2},$ and $\Omega_{0}^{-1},$ respectively (see Eq. \ref{dyn}).
}
\label{figure:leftslab}
\end{figure}

The magnitude of the force acting on the particle
of the cross-section $d^{2}$ can be evaluated with Eq. (\ref{force})
as $P_{0}d^{2},$ that is proportional to the injected into the waveguide light power $W_0$ \cite{Nemoto,Romero}.
At $W_0=100$ mW, this yields the typical optical force $F$ of the order of 1 nN.
The characteristic frequency of the oscillations, which we need for dimensionless equations of motion,
can be estimated {by an order of magnitude in the physical units
as $\Omega_0=\sqrt{F/dm}$.} Since the dielectric particles of our interest with the
size of the order of $10^{-4}$ cm  have masses $m$ of the order of 1 pg,
these oscillations show characteristic frequencies of the order of $2\pi\times 100$ kHz \cite{Romero},
much lower than the light frequency.
{Below we show as dependent on the initial conditions motion of particles can be bounded
with characteristic frequency substantially less than $\Omega_0$ or unbounded on times
considerably larger than $\Omega_0^{-1}$.} {The corresponding velocity of $\Omega_{0}d$ being
of the order of 10 cm/s allows one to consider the light transmission adiabatically. On the other hand,
the thermal velocity of a particle of the mass of 1 pg at the room temperature is of the order of 1 mm/s, which
allows one to a good approximation neglect the random Brownian motion.}

Introducing the dimensionless coordinate via $d,$ force as $P_{0}d^{2}$ and mass $m\equiv 1$ (leading to the  time unit as $\Omega_0^{-1}$),
we can write Eq. (\ref{dynL}) in dimensionless form
\begin{equation}\label{dyn}
    \ddot{x}_j+\gamma\dot{x}_j=f_j(x_j),
\end{equation}
where $f_{j}$ is the dimensionless force acting on the $j$-th particle  \cite{Praveen}, and $\gamma$ is expressed in
the physical units as
\begin{equation}\label{gamma}
    \gamma=6\pi\eta \sqrt{\frac{d}{P_{0}m}}.
\end{equation}
For water with $\eta_{w}\approx 10^{-2}$ $\mbox{dyn}\cdot\mbox{s/cm}^2$, Eq. (\ref{gamma})
yields the dimensionless $\gamma$ of the order of 10 for the injected light 
power $W_{0}$ of the order of 100 mW. For the air with $\eta_{a}\approx 0.01\eta_{w}$, 
the value of $\gamma$ at the same $W_{0}$ is of the order of 0.1, corresponding to a relatively weak damping.
With the increase in the light power, the effect of viscous friction decreases as $W_{0}^{-1/2}.$

For the realization corresponding to Fig. \ref{figure:scatteringcenter}(a), 
we show in Fig. \ref{figure:leftslab}(a) the light transmittance through two particles,
optical force $f_L(x_L)$ and the corresponding potential
\begin{equation}
\label{UL}
U_L(x_L)=-\int_{x_L(0)}^{x_L}f_L(x)dx,
\end{equation}
where $x_L(0)$ is the initial position of the particle.
One can see that in the Newtonian regime $\gamma=0$ {describing exactly particles in the vacuum or approximately in the air,} we have
either the bounded or unbounded time evolution of the positions, dependent on
$x_{L}(0)$ as presented in Figs. \ref{figure:leftslab} (a) and {\ref{figure:leftslab}} (b).
The characteristic period of the potential is determined by the wave vector $k$.
Two particles in the waveguide form a Fabry-Perot resonator (FPR)
structure in which the transmittance $|T|^2$ shows sharp peaks when distance between the slab and the 
scattering center equals integer number of half wave lengths. Then the wave function amplitude $\psi(x)$
inside the resonator is maximal to give rise to resonant behavior of the optical force acting on the
walls of the resonator. Indeed, one can see that the force follows the light transmittance  with sharp
resonant negative dips.
As the result the potential $U_L(x_L)$ in (\ref{UL}) acquires a tilted periodic shape with
the particle dynamics qualitatively different from that in a simple periodic one.

\begin{figure}[htbp]
\centering
\includegraphics*[width=6cm,clip=]{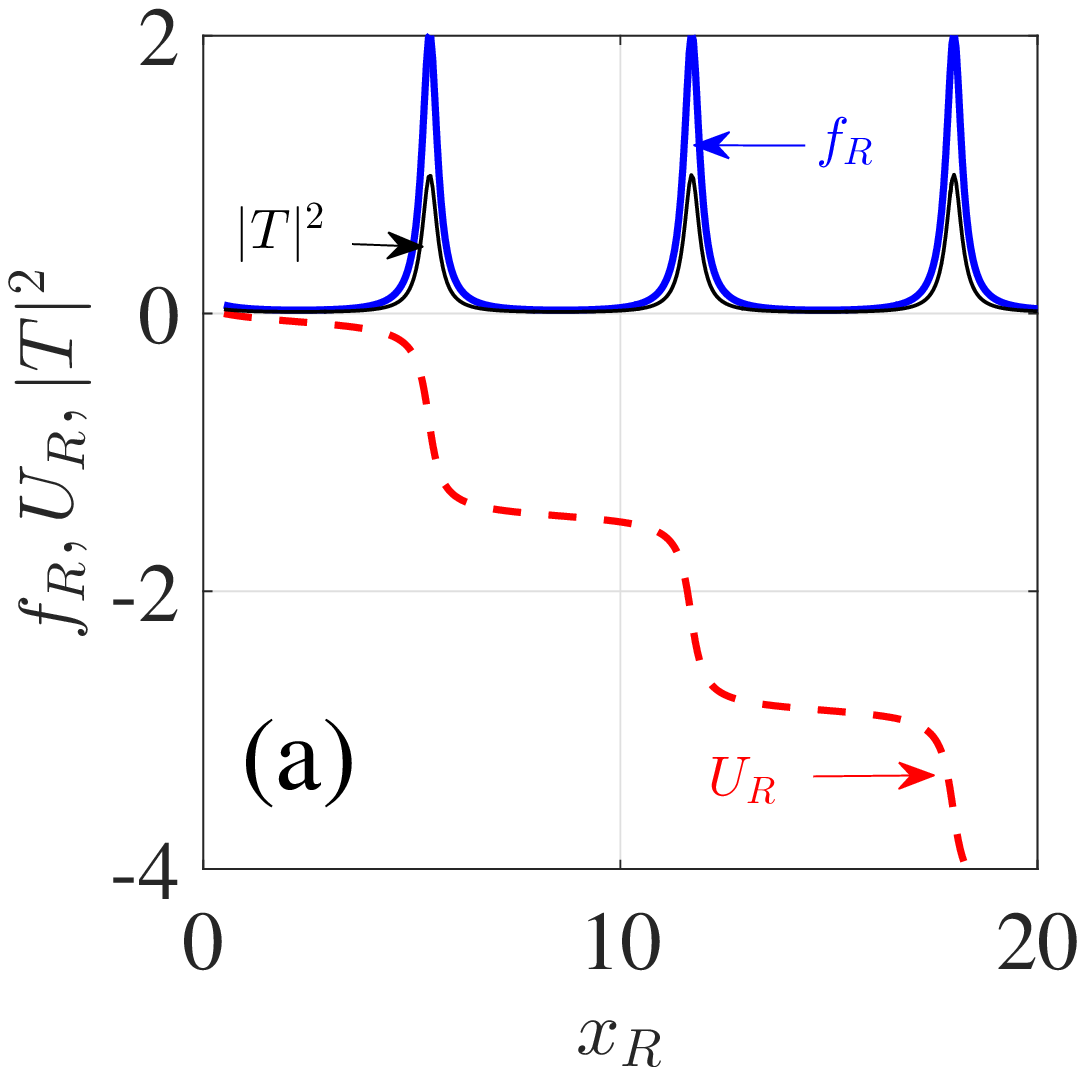}\\
\hspace{-0.5cm}\includegraphics*[width=6cm,clip=]{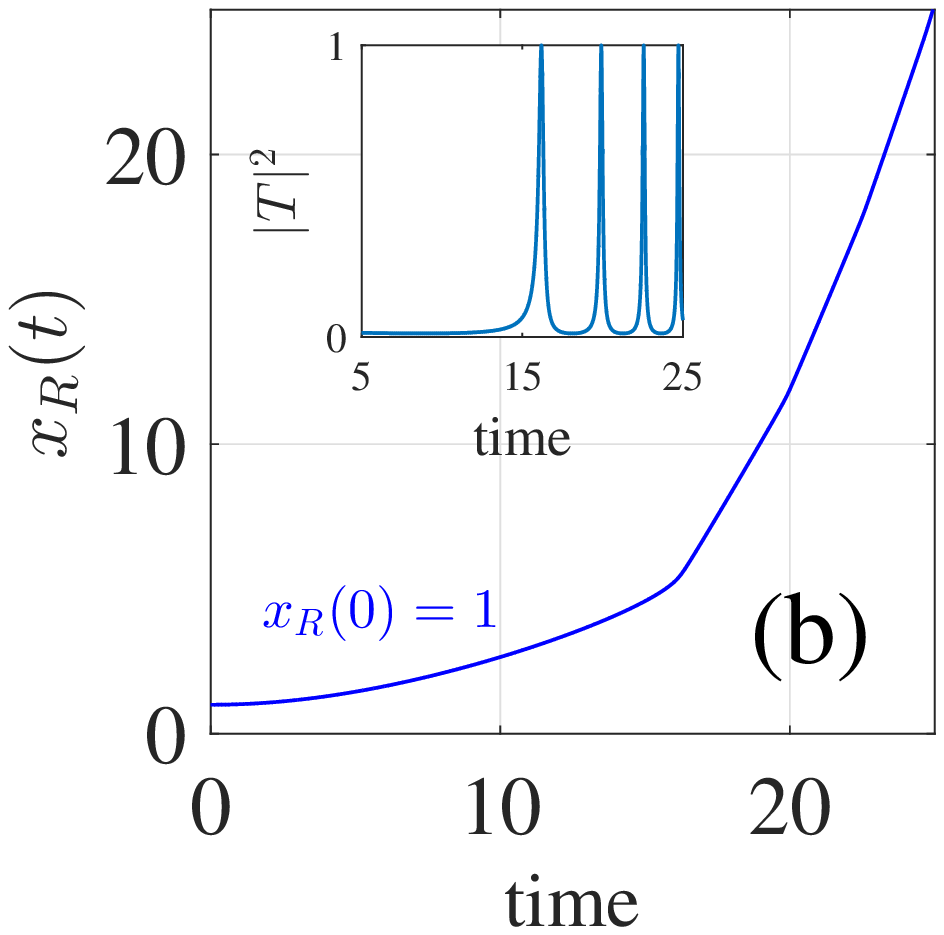}
\caption{(a) The optical force (\ref{forces}) (solid line), the corresponding potential (dot line), and the light transmission (dash-dot line)
for the immobile left particle positioned at $x_{L}=0$. (b) The corresponding
time evolution of the position of the right particle and light transmission (in the inset) for $\gamma=0.$
The parameters of ``scattering center'' and the slab are identical with 
$a=d/2, \epsilon=4$, and $k=1/2$. The coordinates, forces, and time are given in the units of $d$,
$P_{0}d^{2},$ and $\Omega_{0}^{-1},$ respectively (see Eq. \ref{dyn}).}
\label{figure:rightslab}
\end{figure}
Respectively, as depends on the initial
position of the particle, the time evolution shows oscillations
or a motion until the particle touches the immobile element {at $x_{L}=-a$}
as shown in Fig. \ref{figure:leftslab}(b).
{After this event the evolution needs a special analysis
which goes beyond the scope of the present paper.}
The inset in Fig. \ref{figure:leftslab}(b) shows that the choice of the initial position
strongly changes the evolution of the light transmission
through the particles with the left particle dragged by light.
Here we obtain oscillations with growing frequency since
the distance between particles increases with time with acceleration caused by nonzero mean
optical force. For the periodic oscillations
of the left particle shown by the red line in Fig. \ref{figure:leftslab}(b) the time oscillations of light
transmission are periodic with a few harmonics. The appearance of this multi-frequency behavior
is a result of anharmonicity of the binding potential $U_L$ shown in Fig. \ref{figure:leftslab}(a).

\begin{figure}[htbp]
\centering
\includegraphics*[width=6cm,clip=]{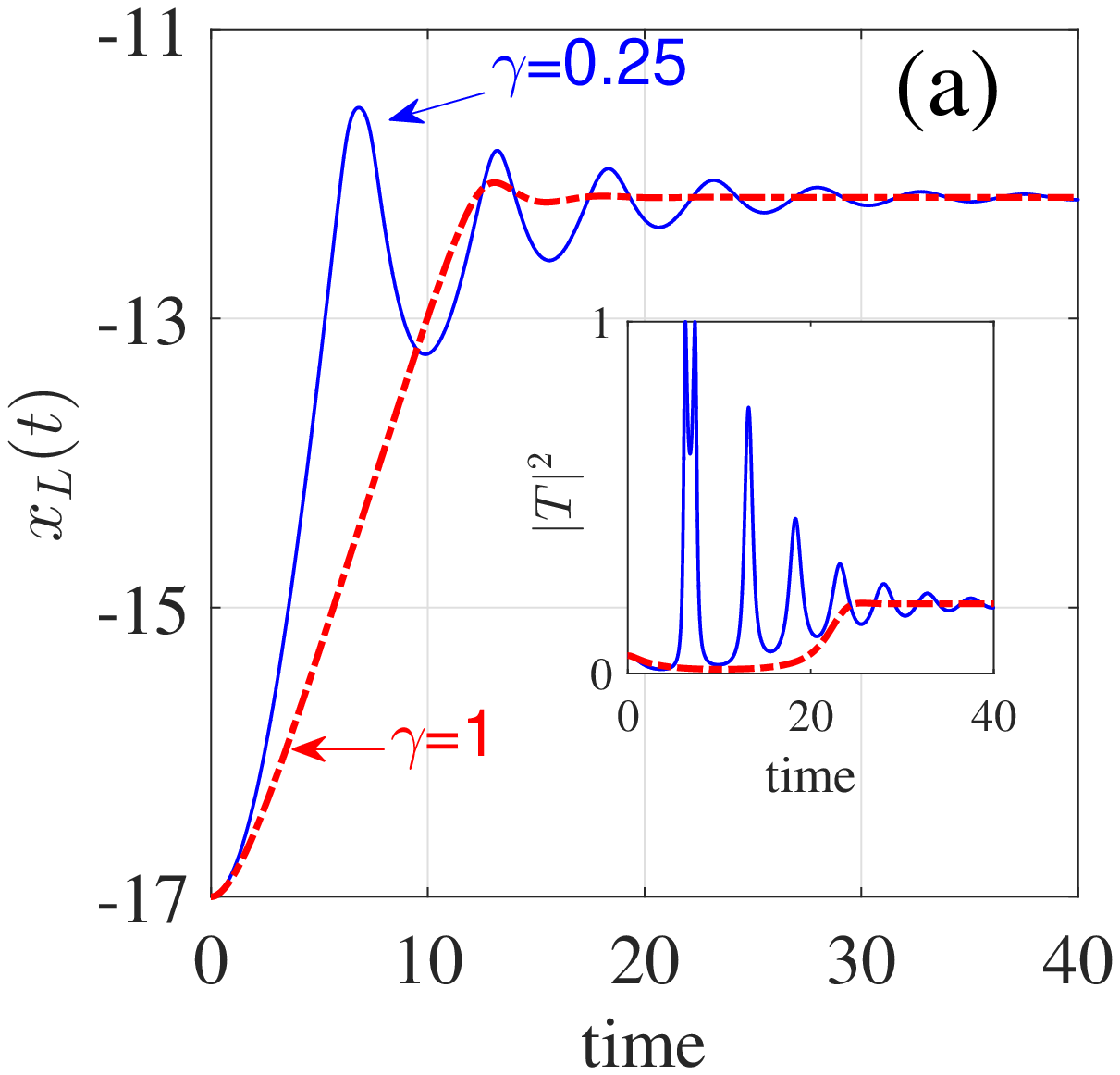}\\
\includegraphics*[width=6cm,clip=]{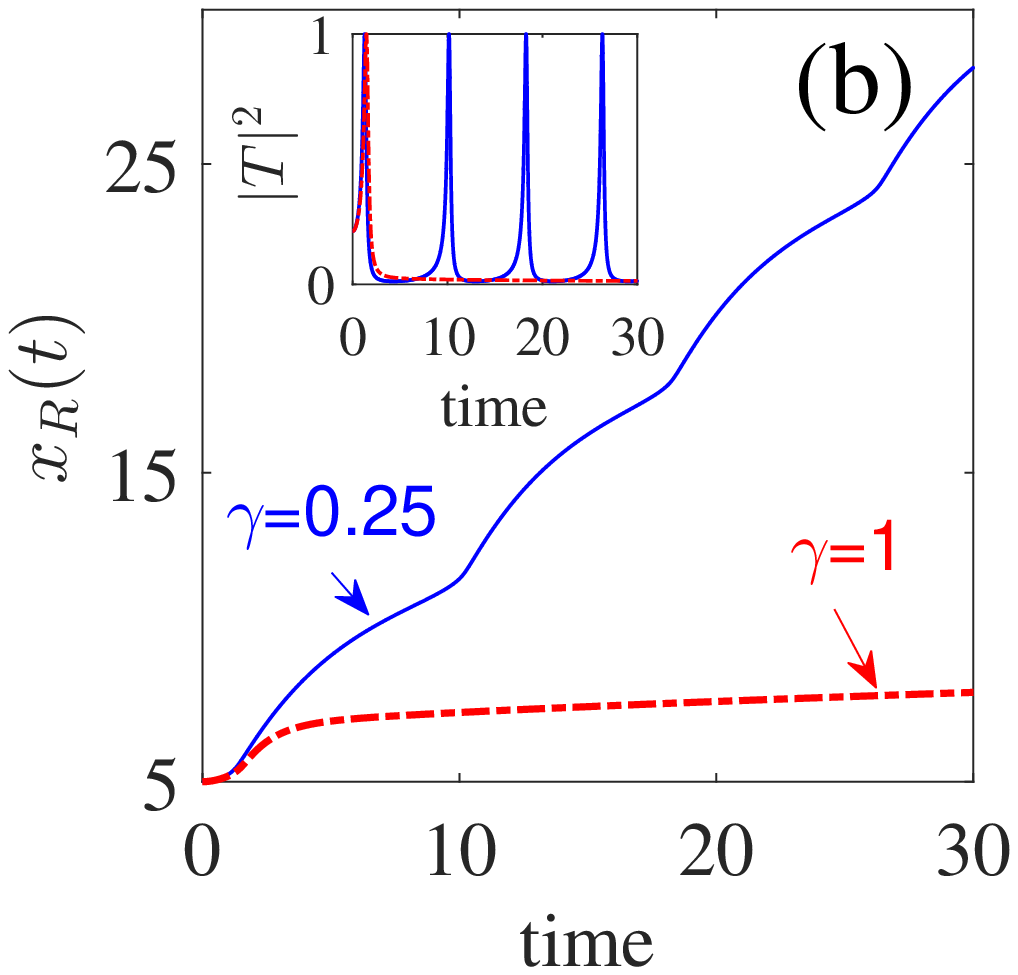}
\caption{Time evolution of the left (a) or right (b) particle coordinate and light transmission in a viscous medium 
with $x_L(0)=-17$ and $x_R(0)=5$. The values of $\gamma$ are shown near the plots. 
The coordinates, forces, and time are given in the units of $d$,
$P_{0}d^{2},$ and $\Omega_{0}^{-1},$ respectively (see Eq. \ref{dyn}).}
\label{figure:viscous}
\end{figure}

Figure \ref{figure:rightslab} shows the case corresponding 
to Fig. \ref{figure:scatteringcenter}(b). Again the optical 
force follows the resonant dependence of the transmittance as a function 
of the distance between the particles with however positive peaks.
Unlike the case of Fig. \ref{figure:scatteringcenter}(a), the motion of the particle here 
is always unbounded. Respectively, we have time oscillations of transmittance in Fig.  \ref{figure:rightslab} 
(b) with growing frequency. Effects of damping on the motion of the particles with the corresponding 
time evolution of light transmittance are shown in Fig. \ref{figure:viscous}.

\section{Evolution in system of two mobile particles}

\begin{figure}
\includegraphics*[width=6cm,clip=]{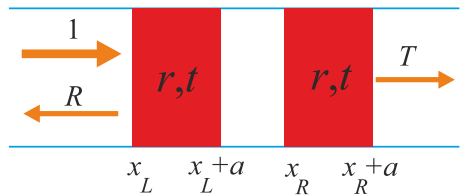}\\
\caption{Two mobile slabs geometry.}
\label{figure:twoslabspicture}
\end{figure}

\begin{figure}
\hspace{-1cm}\includegraphics*[width=5cm,clip=]{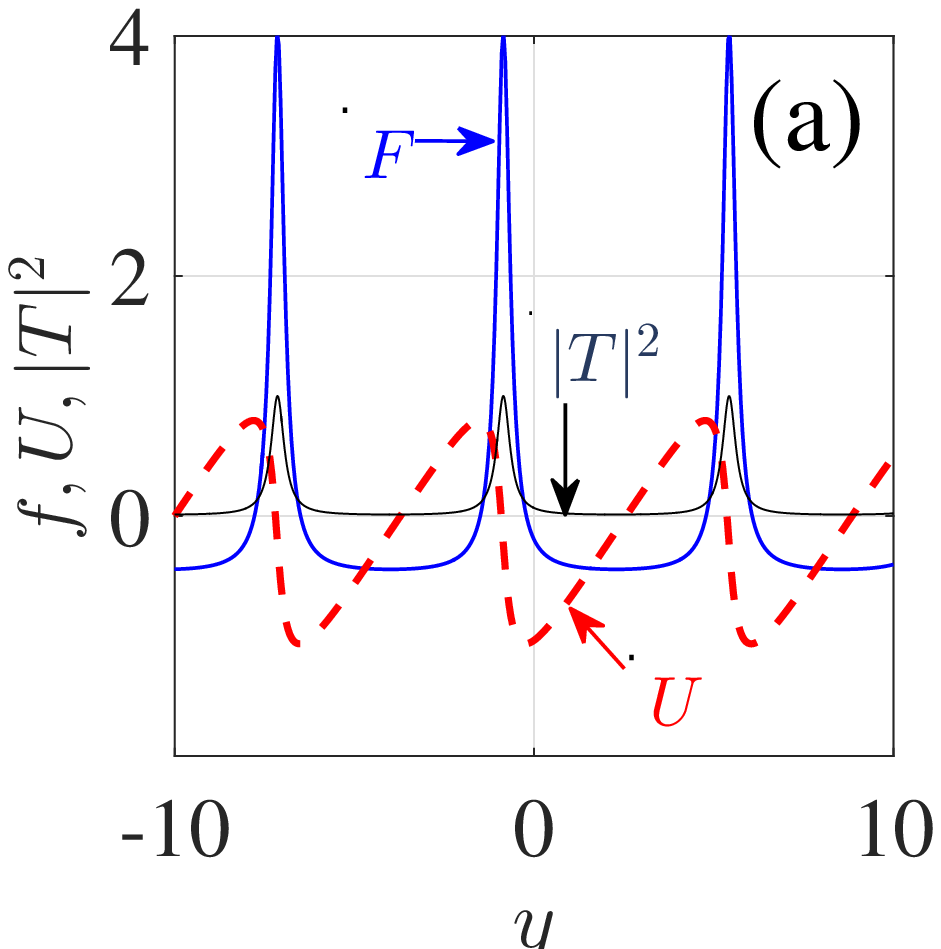}\\
\hspace{-1.3cm}\includegraphics*[width=5.4cm,clip=]{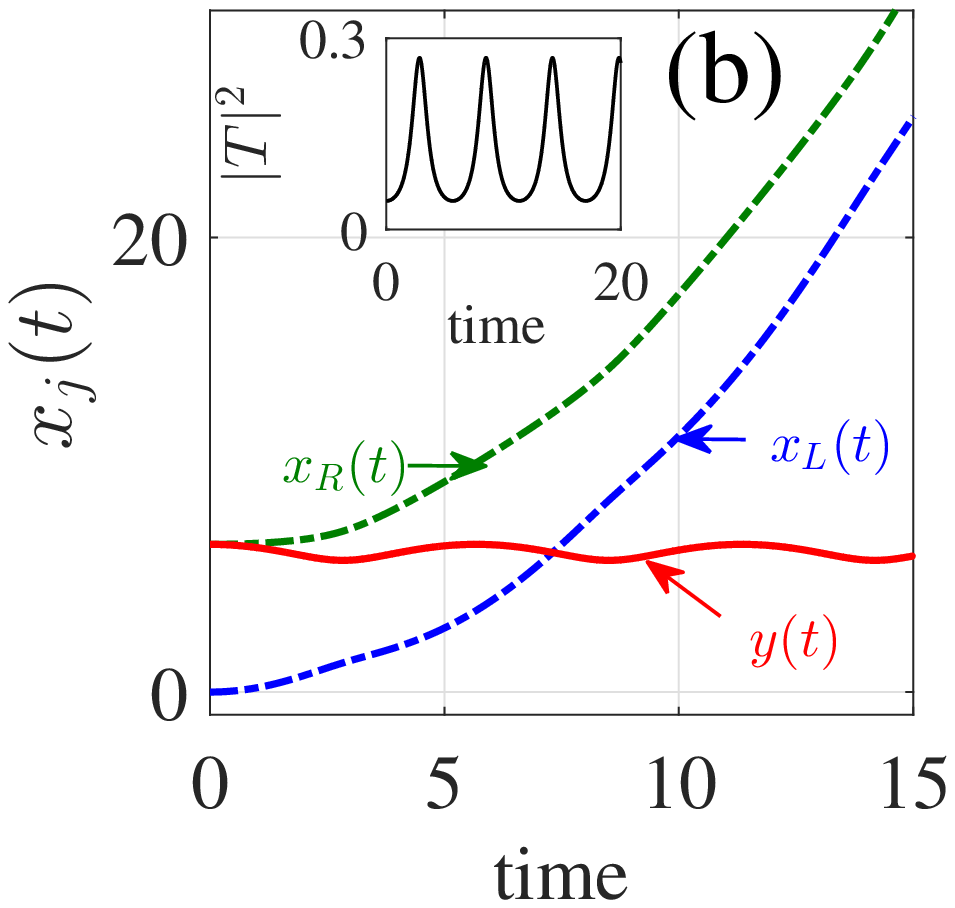}\\
\hspace{-0.3cm}\includegraphics*[width=7.4cm,clip=]{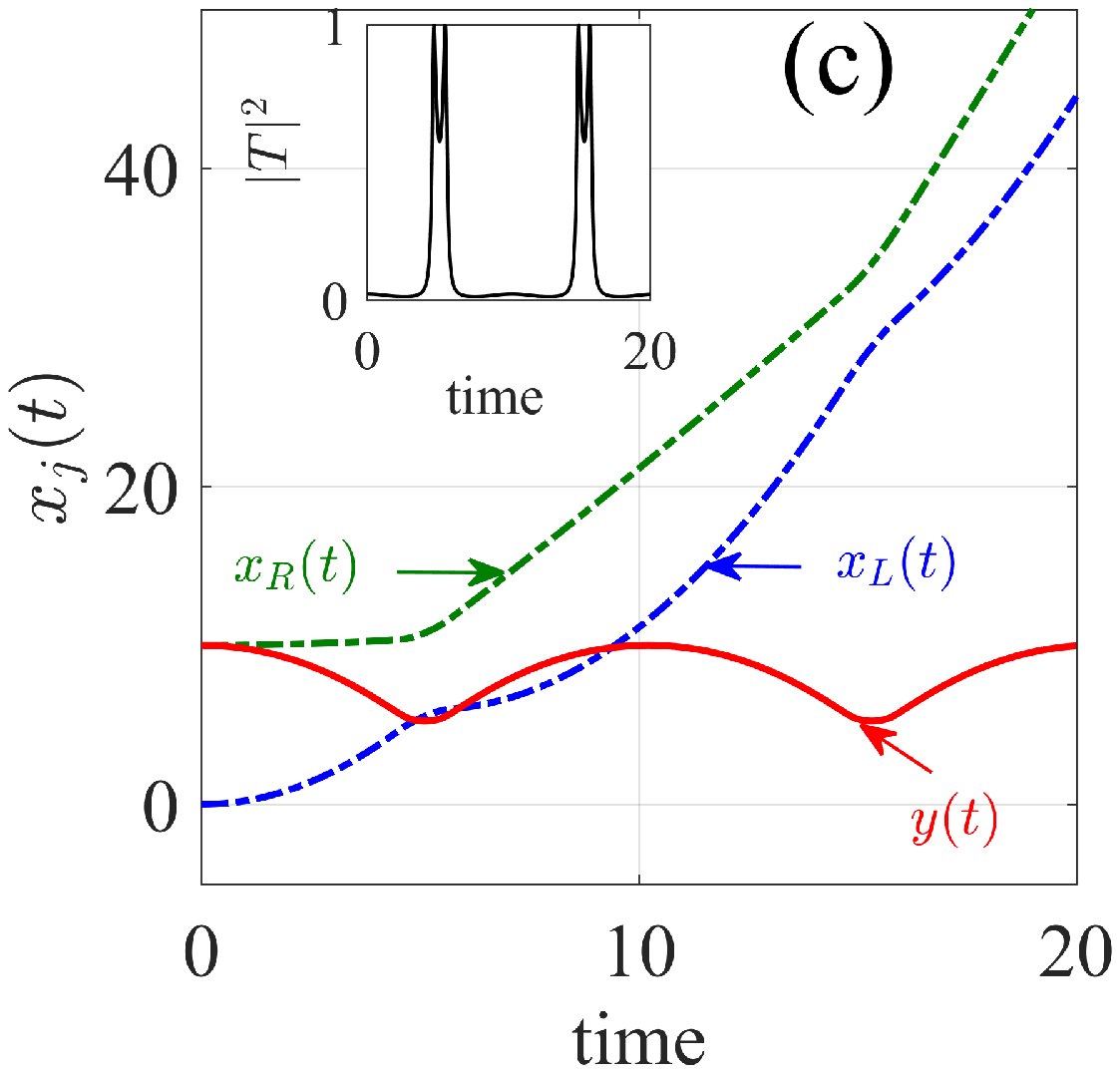}
\caption{ (a) Force and potential vs function of distance
between two mobile particles. Time evolution
of coordinates and distance between the particles for the initial positions
$x_L(0)=0$ and $x_{R}(0)=6.5$ (b) and $x_{R}(0)=10$ (c). The coordinates, forces, and time are given in the units of $d$,
$P_{0}d^{2},$ and $\Omega_{0}^{-1},$ respectively (see Eq. (\ref{dyn})).}
\label{figure:twoslabsmotion}
\end{figure}

For identical particles shown in Fig. \ref{figure:twoslabspicture}  we obtain similarly to
Eqs. (\ref{dynL}) and (\ref{forces}) the following dimensionless equation of motion:
\begin{equation}\label{newton}
    \ddot{y}+\gamma\dot{y}=\widetilde{f}(y)=-\displaystyle{\frac{d\widetilde{U}}{dy}},
\end{equation}
where $\widetilde{f}(y)=f_{R}(y)-f_{L}(y)$.  The ``force'' $\widetilde{f}(y)$ depends only on the distance between particles $y=x_{R}-x_{L}$
and is shown in Fig. \ref{figure:twoslabsmotion}(a). Surprisingly, the corresponding ``potential'' $\widetilde{U}(y)$
shows only periodic dependence on the distance $y,$ different from the interaction
considered in the previous Section and similar to the optical binding of atomic clouds \cite{Asboth} due to the standing
EM waves. The characteristic ``potential'' height $\widetilde{U}_{0}$ can be estimated
as $Fd\sim 10^{-8}\mbox{ erg}$.
Since we consider the light incident from the left, the inversion symmetry is broken resulting in $\widetilde{U}(y)\neq \widetilde{U}(-y)$.

Figure  \ref{figure:twoslabsmotion} demonstrates that the time dependence of the light transmittance
strongly depends on the initial distance between the particles. For the
distance $y(0)$ when the potential is close to the minimum, the positions evolve in time
approximately preserving the interparticle distance. Respectively, the
light transmittance oscillates with time approximately harmonically
as shown in the inset of Fig. \ref{figure:twoslabsmotion}(b).
However if the initial position is far from the minimum,
the nonparabolicity of the potential becomes important and the
time dependence of the transmittance acquires higher harmonics.
Interactions presented in Fig. \ref{figure:twoslabsmotion} show more variety than the
optical binding of small dielectric particles  in the Bessel beams \cite{Karasek} and in the random
fields \cite{Brugger}.

\section{Dynamics of a particle inside a Fabry-Perot resonator}
\begin{figure}
\includegraphics*[width=8cm,clip=]{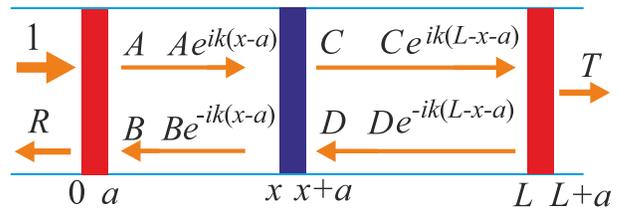}
\caption{Two slabs with dielectric constant $\epsilon_{m}$ shown by red color
fixed at $x=0$ and $x=L$ form FPR (cavity).
The third slab with the dielectric constant $\epsilon$ can move inside the resonator.}
\label{figure:FPRpicture}
\end{figure}

The analysis of the system of dielectric slabs in the waveguide allows us to consider analytically
optical driving of a dielectric particle by EM fields in resonant cavities.
The cavity can be modeled by two immobile dielectric slabs and the particle
is modeled by a mobile slab as shown in Fig. \ref{figure:FPRpicture}.

\begin{figure}
\includegraphics*[width=8cm,clip=]{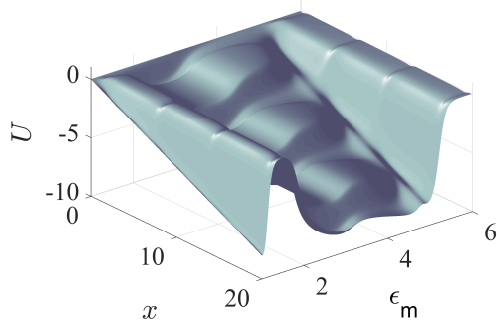}
\caption{The potential $U(x)$ (in the units of $P_{0}d^{3}$) vs the particle position (in the units of $d$) 
and the dielectric constant of
the FPR walls for $\epsilon=4, a=1/2, k=1/2$, and $L=20$.}
\label{figure:FPRpotential}
\end{figure}

The solutions of the EM field equations are given by the transfer matrices
\begin{equation}
\label{ABC}
\left[\begin{array}{c}
  A\\ B\end{array}\right]=\mathbf{m}\left[\begin{array}{c}1\\  R\end{array}\right],
\quad \left[\begin{array}{c}T\\0\end{array}\right]=\mathbf{m}\left[\begin{array}{c}
 Ce^{ik(L-x-a)}\\ De^{-ik(L-x-a)} \end{array}\right]
\end{equation}
for the walls of the resonator and 
\begin{equation}
 \left[\begin{array}{c} C\\ D  \end{array}\right]=\mathbf{M}
  \left[\begin{array}{c}Ae^{ik(x-a)}\\ Be^{-ik(x-a)}\end{array}\right]
\end{equation}
for the embedded movable slab. Here the matrix $\mathbf{M}$ is given by Eq. (\ref{M}), $\mathbf{m}$ has the elements
\begin{eqnarray}\label{m}
m_{11}&=&\cos(q_0a)+\displaystyle{\frac{i}{2}\left[\frac{q_m}{k}+\frac{k}{q_m}\right]
\sin(q_ma)},\\
m_{12}&=&\displaystyle{\frac{i}{2}\left[\frac{q_m}{k}-\frac{k}{q_m}\right]\sin(q_ma)},~
m_{22}=m_{11}^{*}, ~m_{21}=m_{12}^{*},\nonumber
\end{eqnarray}
and
\begin{equation}\label{q0}
 q_m^2=\epsilon_m k^2+(\epsilon_m-1)\frac{\pi^{2}}{d^{2}}.
\end{equation}
Similar to Eq. (\ref{force1}) we have for the optical pressure on the mobile particle
\cite{Asboth}
\begin{equation}\label{forceP}
    P(x)=P_0\left[|A(x)|^2+|B(x)|^2-|C(x)|^2-|D(x)|^2\right].
\end{equation}
The corresponding potential $U(x)$ is presented in Fig. \ref{figure:FPRpotential}. One can see
that it strongly depends on the dielectric constant of the walls of the resonator,
i.e. on its openness. For $\epsilon_m$ close to $\epsilon$ the potential holds
local minima capable to bind the particle at the corresponding positions. This result is reminiscence
of electron transmission through the well potential relief with two different potential wells \cite{Ricco}.

\begin{figure}
\hspace{-0.51cm}\includegraphics*[width=8cm,clip=]{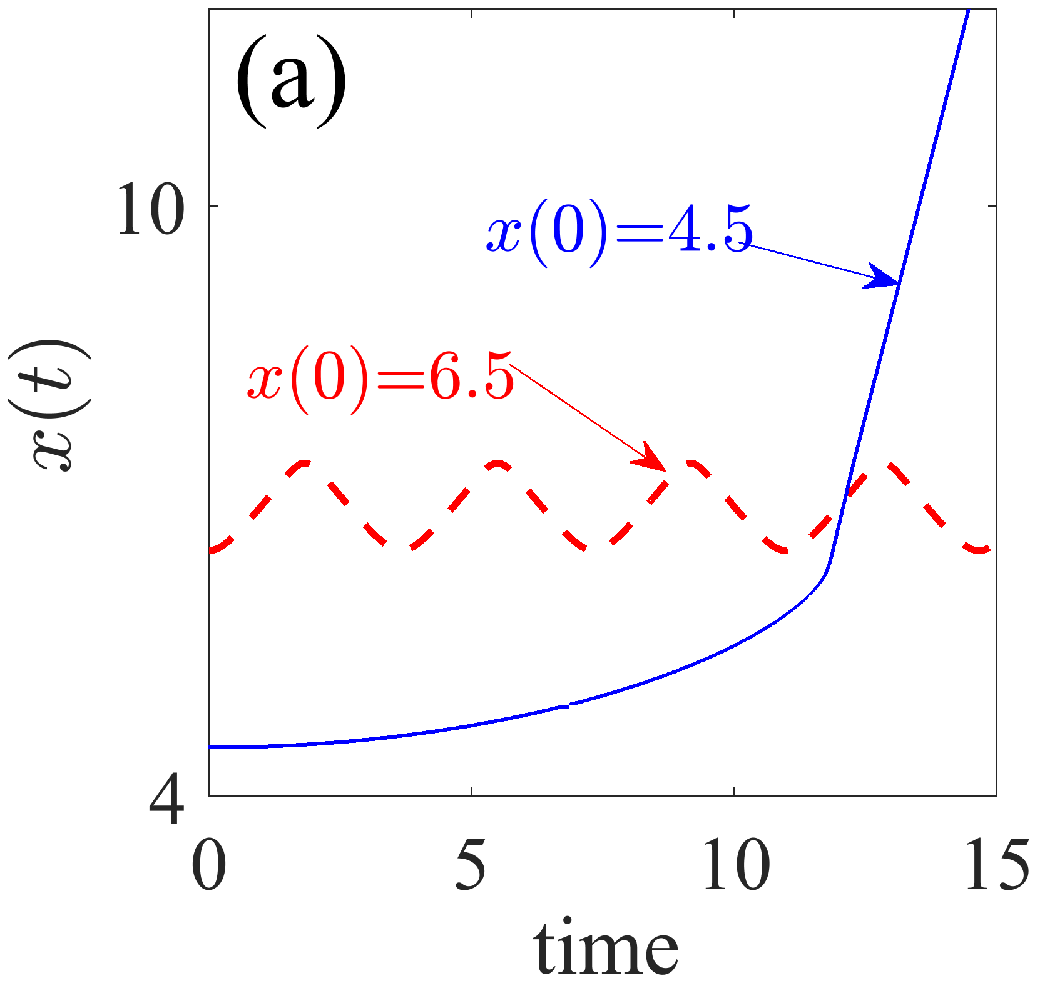}
\includegraphics*[width=9cm,clip=]{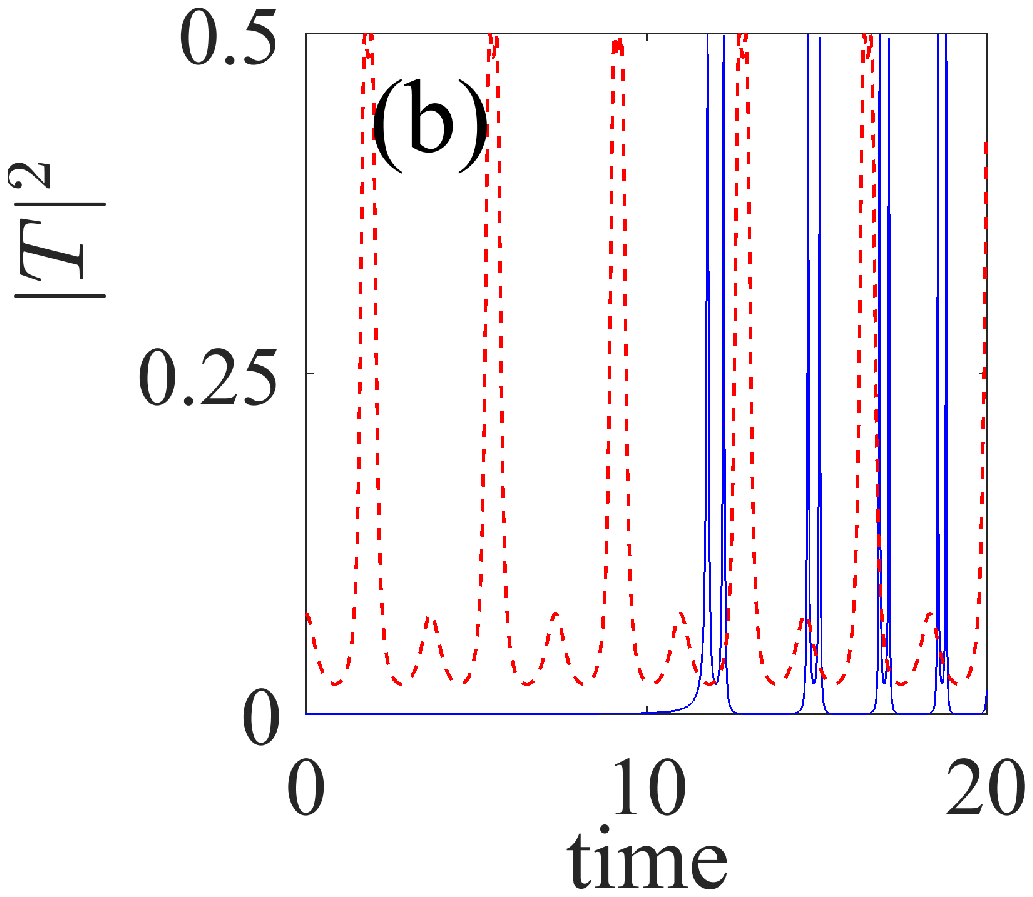}
\caption{ Time evolution of the particle position (a) and transmittance (b) for
the same parameters
given in Fig. \ref{figure:FPRpotential}, $\epsilon_{m}=3$ and $L=20$. The initial position of the particle
is $x(0)=6.5$ (red dash-dot lines) and  $x(0)=4.5$ (blue solid lines). 
The coordinates are given in the units of $d.$}
\label{figure:FPRmotion}
\end{figure}

Time evolution of the particle position in the FPR is shown in
Fig. \ref{figure:FPRmotion}(a) for two initial $x(0).$ The first position, $x(0)=6.5,$ yields oscillations
in the vicinity of a local potential minimum shown in Fig. \ref{figure:FPRpotential}. An extremely nonlinear
profile of the potential over the particle position gives rise to the shape of the corresponding time
oscillations of the light transmittance shown in Fig. \ref{figure:FPRmotion}(b) by dashed red line. The second
choice, $x(0)=4.5,$ corresponds to the accelerated time evolution until the particle will reach
the right wall. This motion corresponds to the time behavior of
the transmittance as shown in Fig. \ref{figure:FPRmotion}(b) by blue solid line.

\section{Summary and conclusions}

Although the replacement of the particles by slabs
as   shown in Fig. \ref{figure:model} is a significant simplification, it preserves the main qualitative
features of the light transmittance in a waveguide with embedded particles {as described in general terms
by a transfer-matrix dependent on the positions and optical properties of these particles.}
When one particle is inserted in a waveguide, it is subject to a radiation pressure
of the propagating light. This pressure does not depend on the position
of the particle and produces its constant acceleration in the vacuum or drags it
{in a viscous medium} with a constant velocity. The
transmittance of light through the particle remains {position and time-independent.} The situation
changes dramatically if at least two particles are inserted in the waveguide. Because of
different light pressure acting on the particles the interparticle distance
changes with time. Respectively, the transmittance given by the Fabry-Perot resonator
{transfer matrix equations} (\ref{RT}) acquires a
time dependence. 

To describe the light-induced interaction between the particles one
can introduce an effective system-dependent potential. This potential
usually has a tilted (or a simple, as depends on the system realization) periodic shape, where
the evolution of the interparticle distance can be bounded or unbounded.
As a result, the light transmittance shows a rich variety of time-dependent behaviors in the form of
time oscillations either with a few harmonics {for a bounded motion or with a growing frequency
for the unbounded one.}
The characteristic period of the oscillations {in the light transmittance} shown in the paper is
of the order of $10^{-5}$ s for the propagating light power of the order of
100 mW. Therefore, by changing the laser light power and direction one can achieve different regimes of the particles motion. 
Similar modifications can be achieved by choosing different materials for the movable particles 
and static elements such as the ``scattering center'' in Fig. \ref{figure:scatteringcenter} or walls of the Fabry-Perot resonator
in Fig. \ref{figure:FPRpicture}.

It is important to mention that recent publications confirm the presence of this phenomenon
in  different experimental set-ups: two rotating dielectric
microparticles \cite{Arita} and density oscillations of swimming
bacteria confined in microchambers \cite{Paoluzzi}. Both systems show the
characteristic frequencies of light modulation in the sound range.
Thus, the analysis of the time dependence of the light transmittance paves a way
for manipulating and monitoring motion of the particles in optical waveguides.

\begin{acknowledgments} The work of A.F.S. was  partially  supported  by grant 
14-12-00266 from the Russian Science Foundation. This work of E.Y.S was supported by the
University of Basque Country UPV/EHU under program UFI 11/55, FIS2015-67161-P (MINECO/FEDER), 
and Grupos Consolidados UPV/EHU del Gobierno Vasco (IT-472-10).
\end{acknowledgments}


\begin{thebibliography}{99}
\bibitem{coefficient}
The numerical coefficient in the viscosity force expression depends on the shape of the partice with $6\pi$ being, technically speaking,
correct only for a perfect sphere. Nevertheless, we will use it as an order-of-magnitude estimate.
\end{thebibliography}
\end{document}